\documentclass{svmult}

\usepackage{mathptmx}       % selects Times Roman as basic font
\usepackage{helvet}         % selects Helvetica as sans-serif font
\usepackage{courier}        % selects Courier as typewriter font
\usepackage{makeidx}         % allows index generation
\usepackage{graphicx}        % standard LaTeX graphics tool
\usepackage{multicol}        % used for the two-column index
\usepackage[bottom]{footmisc}% places footnotes at page bottom

\usepackage[utf8]{inputenc}
\usepackage[T1]{fontenc}
\usepackage{amsmath}
\usepackage{amssymb}
\usepackage{mathrsfs}

\usepackage{microtype}
\usepackage{cite}

\newcommand*{\scri}{\mathscr{I}}

\newcommand*{\del}{\partial}

\begin{document}

\title*{The Spin-2 Equation on Minkowski Background}
% Use \titlerunning{Short Title} for an abbreviated version of your contribution title if
% the original one is too long
\author{Florian Beyer \and George Doulis \and J\"org Frauendiener \and Ben Whale}
% Use \authorrunning{Short Title} for an abbreviated version of your contribution title if
% the original one is too long \institute{Florian Beyer \email{fbeyer@maths.otago.ac.nz}
% Florian Beyer \email{fbeyer@maths.otago.ac.nz} \at Department of Mathematics and
% Statistics, University of Otago, P.O. Box 56, Dunedin 9010, New Zealand
\institute{Florian Beyer \at Department of Mathematics and Statistics, University of
  Otago, P.O. Box 56, Dunedin 9010, New Zealand \email{fbeyer@maths.otago.ac.nz} \and
  George Doulis \at Department of Mathematics and Statistics, University of Otago,
  P.O. Box 56, Dunedin 9010, New Zealand \email{gdoulis@maths.otago.ac.nz} \and J\"org
  Frauendiener \at Department of Mathematics and Statistics, University of Otago, P.O. Box
  56, Dunedin 9010, New Zealand \email{joergf@maths.otago.ac.nz} \and Ben Whale \at
  Department of Mathematics and Statistics, University of Otago, P.O. Box 56, Dunedin
  9010, New Zealand \email{bwhale@maths.otago.ac.nz} }
%
% Use the package "url.sty" to avoid problems with special characters used in your e-mail
% or web address
%
\maketitle

\abstract{The linearised general conformal field equations in their first and
second order form are used to study the behaviour of the spin-2 zero-rest-mass equation on Minkowski background in the vicinity of space-like infinity.}

\section{Introduction}
\label{sec:introduction}

In~\cite{Friedrich:1998tc}, a completely novel finite representation of space-like
infinity was proposed. In this setting, space-like infinity $i^0$ ``blows up'' to
a cylinder $I$ and, consequently, asymptotically flat initial data become regular near $i^0$ and
can be prescribed on generic space-like Cauchy surfaces, 
%i.e. the use of 
%hyperboloidal hypersurfaces is not essential anymore~\cite{Moncrief:2009ds}. 
%% I would not phrase it so bluntly. Maybe more like
i.e., the use of hyperboloidal hypersurfaces may ultimately be 
unnecessary~\cite{Moncrief:2009ds}.
It turns out that, in this representation, Friedrich’s 
general conformal field equations (GCFE) acquire a very simple form and the 
cylinder becomes a total characteristic of the system in the sense that there are no radial derivatives 
in the equations restricted to $I$. In addition, the coordinate location of null infinity is now known beforehand and it does not have to be determined during the evolution.
All these features, 
namely the regularity of the initial data, 
%the fixed location of null infinity, 
the a-priori fixed finite coordinate location of null infinity,
the fact that in the Minkowski case the entire physical space-time can be
covered by one computational domain, 
%%Florian: This is not necessarily really true since we still may need more than one coordinate patch, for example to cover the angular part. I think that what you mean is best expressed by adding the word "finite" to the previous statement.
and the extremely simple form of the 
evolution equations (especially on the cylinder), make the general conformal 
field equations suitable for numerical manipulations. 
However, as expected, the intrinsic system of evolution equations on 
the cylinder degenerates at the interface of the cylinder $I$ with null infinity.
In general, the solutions generate logarithmic singularities at these regions
%which will travel along null infinity and spoil its smoothness, making it 
which are expected to travel along null infinity and spoil its smoothness, making it 
impossible to read-off the gravitational radiation at $\scri^+$. 
%\textbf{Florian}: You make it sound like that this ``drawback'' is a consequence of %Helmut's representation of the equations. On the contrary, it is a general fact about %Einstein's equations that there are asymptotically flat data which do NOT evolve into %a solution with a smooth null infinity. And it one of the *successes* of Helmut's %approach to unveil this surprising and important result! Clearly, this fact could not %be derived easily when one considers, for example, the one-point compactification of %spacelike infinity.
The way 
out of this problem is to prescribe initial data that respect the regularity
conditions proposed in~\cite{Friedrich:1998tc}. 

In this short contribution, we use Friedrich's general conformal field 
equations to evolve generic asymptotically flat
%\textbf{Florian}: We need to decide if we want to call these data ``asymptotically %flat'' or ``asymptotically Euclidean'' and then stick to it. I would be happy with %both
initial data near space-like infinity. We begin our endeavour from the simplest
possible case: linearised gravitational fields on a Minkowski background. 
Although simple, this ``toy model'' encapsulates most of the crucial characteristics
of the full non-linear system described above. 
%Namely, the cylinder must be introduced in order 
%to set our initial data regular and the intrinsic to the cylinder evolution
%equations degenerate at the regions $I^\pm$ that the cylinder meets null 
%infinity. 
The initial data are evolved as close as possible to the 
ill-behaved regions $I^\pm$ and study the behaviour of the numerical solutions
there. This procedure is carried out twice by using the linearised general
conformal field equations in their first and second order form. Analytically the
two approaches are equivalent, but their numerical implementation could very 
well differ. The latter statement is partly based on the claim made in
\cite{Kreiss:2002du} that writing the equations of general relativity as a system
of second order PDE's is more advantageous numerically than writing them as a
system of first order PDE's. According to \cite{Kreiss:2002du}, numerical
simulations based on second order PDE's have better numerical accuracy and avoid
the appearance of spurious waves travelling against the characteristic curves.

\section{The spin-2 equations}
\label{sec:spin-2-zero}

We use the spin-2 zero-rest-mass equation for a totally symmetric spinor field
$\phi_{ABCD}$ in the 2-spinor formulation, i.e. $\nabla_{A'}{}^A \phi_{ABCD} = 0$, 
to model linearised gravitational fields on a Minkowski background. Taking the
components of the above expression, decomposing the five independent components
$\phi_k$ of the spin-2 into harmonic modes $(l,m)$, 
%\textbf{Florian}: You can always decompose the functions in terms of ``harmonic %modes'' even when you don't have spherical symmetry. What is special about spherical %symmetry, however, is that the evolution equations for these modes decouple and hence %we can always assume a fixed value of $l$ and $m$ below ... 
introducing coordinates $(t,r,\theta,\varphi)$ and an adapted spin-frame on the
cylinder, the above spin-2 equation splits into the
eight coupled equations (for details see \cite{Beyer:2012ie})
\begin{equation}
   \label{eq:1}
   \begin{aligned}
     (1-t\kappa') \del_t \phi_k + \kappa \del_r \phi_k - (3\kappa'  - (5-k) \mu)
      \phi_k &=  \mu\, c_k\, \phi_{k-1} , &k=1:4,\\
     (1+t\kappa') \del_t \phi_{k} - \kappa \del_r \phi_{k} + (3\kappa' + (k+1) 
     \mu ) \phi_{k} &= -\mu\, c_k\, \phi_{k+1}, &k=0:3,
   \end{aligned}
 \end{equation}
where $\kappa(r)=r\, \mu(r)$, $\mu(0)=1$, $c_k = \sqrt{l(l + 1) - (2-k)(1-k)}$, 
and $\,'$ denotes differentiation with respect to $r$. Equations \eqref{eq:1}, 
when appropriately combined, split into three constraint equations and a symmetric
hyperbolic (in the domain $|t|<\kappa^{-1}$) system of five evolution equations, 
see \cite{Beyer:2012ie}.

By differentiating the above spin-2 equation, one 
can derive (see \cite{Dou&Frau:2013}) the second order spin-2 wave equation 
 $\square\, \phi_{ABCD} = 0$. Decomposing again the
components of the spin-2 field into harmonic modes and introducing coordinates
$(t,r,\theta,\varphi)$, the spin-2 wave equation splits into a
hyperbolic (in the domain $|t|<\kappa^{-1}$) system of five wave equations
\begin{equation}
 \label{eq:2}
  \begin{aligned}
    (1 - t^2 \kappa'^2) \partial_{tt}\phi_k &- \kappa^2 \partial_{rr}\phi_k + 2 \, t \kappa \kappa' \partial_{tr}\phi_k + 2 \, r \kappa \mu' \partial_{r}\phi_k\\
    &+ 2 \, [(2 - k) \kappa' - t (\kappa'^2 +
    r\, \mu' \kappa' - \frac{1}{2} \kappa \, \kappa'')] \partial_{t}\phi_k  \\
    &+ [(2 - k) (\kappa \kappa'' + (1 - k) \kappa'^2) +k (5 - k) r^2 \mu'^2] \phi_k \\
    & \hspace*{3em}
    = - \mu^2 \, c^2_k \phi_k - r\, \mu \, \mu' (
   (4 - k) c_k \,  \phi_{k+1} +  k \, c_{k-1} \, \phi_{k-1}).
  \end{aligned} 
\end{equation}
%% Here one could unify the notations for \alpha_x and c_k, which saves one definition. I would eliminate the \alpha's and replace them with c_k
Notice that the domain of
 hyperbolicity for both sets of equations is the same, namely $|t|<\kappa^{-1}$. 
Thus, as expected, in both approaches the equations degenerate at 
$(t,r)=(\pm 1,0)$, i.e. at the regions $I^\pm$ where null like infinity 
$\mathscr{I}^\pm$ meets the cylinder. 

In order to compute the same solution as with the first order system \eqref{eq:1}
we have to use all the available
information from the first-order system to determine initial and boundary
conditions for the second order system \eqref{eq:2}. In \cite{Dou&Frau:2013} 
% a correspondence between the two systems has been established that ensures 
% that the first order system holds on both the initial hypersurface and the
% boundaries of the second order system. 
it was shown that the two formulations are equivalent provided the initial and boundary data for the second order system are determined from the first order system.

\section{Numerical results}
\label{sec:numerics}

The PDE systems \eqref{eq:1}, \eqref{eq:2} are discretised according to 
the method of lines. An equidistant grid on the computational domain $D=[0,1]$
is used to discretise the spatial coordinate $r$. The spatial derivatives are approximated by summation by parts (SBP) finite difference operators 
\cite{Mattsson:2004br}. 
%%%%%%%%%%%%%%%%%%%%%%%%%%%%%%%%%%%  table  %%%%%%%%%%%%%%%%%%%%%%%%%%%%%%%%%%%%%%%%%
\begin{table}[htb]
 \centering
  \begin{tabular}{|c|cc|ccc|}
    \hline
          &    \qquad \qquad $\phi_0$        & &   \qquad\qquad\qquad$\phi_4$ &&\\
    \hline
    Grid  &      1st order     &   2nd order  &    1st order     & 2nd order   &\\
    \hline
    50    &     -25.2218       &   -27.6006   &    -11.1643      &  -12.3418   &\\
    \hline 
    100   &     -29.3956       &   -31.5941   &    -13.9924      &  -15.1743   &\\ 
    \hline  
    200   &     -33.7109       &   -35.6075   &    -16.9978      &  -18.1782   &\\ 
    \hline
    400   &     -38.1000       &   -39.6068   &    -20.1075      &  -21.2850   &\\
    \hline
  \end{tabular}
  \caption{The logarithm of the normalized $l^2$ norm of the absolute error $E$,
    $\log_2(||E||_2)$, between the exact solution and the solutions computed from the
    1st-order system~\eqref{eq:1} and the 2nd-order system~\eqref{eq:2} at
    time $t=1$. 
  }
  \label{tab:compar_exact}
\end{table}
%%%%%%%%%%%%%%%%%%%%%%%%%%%%%%%%%%%%%%%%%%%%%%%%%%%%%%%%%%%%%%%%%%%%%%%%%%%%%%%%%%%%%%
The boundary conditions are implemented with a very
simple, but highly efficient, simultaneous approximation term (SAT) penalty method
\cite{Carpenter1994TimeStable}. The temporal integration is based on a standard
explicit fourth order Runge-Kutta scheme. The code has been written form scratch 
in Python. 

The comparison of the numerical properties of the two approaches is based on 
their ability to reproduce a specific family of exact solutions of the spin-2 
equation developed in \cite{Beyer:2012ie}. As was shown in \cite{Beyer:2012ie,Dou&Frau:2013}, the critical sets $I^+$, located at $t=1$, can be reached in both approaches without
loss of the expected 4th order accuracy; in addition, the constraint quantities are
preserved to sufficient accuracy during the evolution. A comparison of the accuracy with which the two 
approaches numerically reproduce the exact solution is shown in  Tab.~\ref{tab:compar_exact}. Better accuracy was achieved in the second order 
formulation, a result that confirms the first claim in \cite{Kreiss:2002du}. The 
second claim made therein that the spurious waves disappear in the second order 
case is also confirmed as the high frequency features disappear in the convergence 
 plots of the second order formulation, see \cite{Dou&Frau:2013}. 

\section{Conclusion}
\label{sec:conclusion}

In this work, two distinct approaches to the linearised general conformal field
equations were developed and subsequently implemented numerically. In both approaches
we managed to reach without loss of accuracy the ill-behaved region $I^+$. It  
is principally not possible to go beyond $I^\pm$ since the equations loose hyperbolicity. A possible way to resolve this problem is presented in \cite{Beyer:2013}. 
We have also shown that the second order formulation of the spin-2 equation leads to a
better accuracy by a factor of 3-4 and the spurious waves travelling against the characteristics disappear, confirming the claims made in \cite{Kreiss:2002du}.

% \bibliographystyle{spmpsci} 
% \bibliography{papers}

\begin{thebibliography}{10}
\providecommand{\url}[1]{#1}
\providecommand{\urlprefix}{URL }
\expandafter\ifx\csname urlstyle\endcsname\relax
  \providecommand{\doi}[1]{DOI~\discretionary{}{}{}#1}\else
  \providecommand{\doi}{DOI~\discretionary{}{}{}\begingroup
  \urlstyle{rm}\Url}\fi
  
\bibitem{Beyer:2012ie}
Beyer, F., Doulis, G., Frauendiener, J., Whale, B.: {Numerical space-times near
space-like  and null infinity. The spin-2 system on Minkowski space}.
\newblock Class. Quantum Grav. \textbf{29}, 245013 (2012)

\bibitem{Beyer:2013}
Beyer, F., Doulis, G., Frauendiener, J., Whale, B.: {Linearized gravitational waves 
near space-like and null infinity}.
\newblock (this volume), arXiv:1302.0043 (2013)

\bibitem{Carpenter1994TimeStable}
Carpenter, M.H., Gottlieb, D., Abarbanel, S.: {Time-stable boundary conditions
  for finite-difference schemes solving hyperbolic systems: methodology and
  application to high-order compact schemes}.
\newblock J. Comp. Phys. \textbf{111}, 220--236 (1994)

\bibitem{Dou&Frau:2013} 
Doulis, G., Frauendiener, J.: {The second order spin-2 system in flat space near
space-like and null-infinity}.  
\newblock to appear in Gen. Relativ. Gravit., arXiv:1301.4286 (2013)

\bibitem{Friedrich:1998tc}
Friedrich, H.: {Gravitational fields near space-like and null infinity}.
\newblock J. Geom. Phys. \textbf{24}, 83--163 (1998)

\bibitem{Kreiss:2002du}
Kreiss, H.O., Ortiz, O.E.: {Some mathematical and numerical questions connected
  with first and second order time-dependent systems of partial differential
  equations}.
\newblock Lect. Notes Phys. 604, 359--370 (2002)

\bibitem{Mattsson:2004br}
Mattsson, K., Nordstr{\"o}m, J.: {Summation by parts operators for finite
  difference approximations of second derivatives}.
\newblock J. Comp. Phys. \textbf{199}, 503--540 (2004)

\bibitem{Moncrief:2009ds}
Moncrief, V., Rinne, O.: {Regularity of the Einstein equations at future 
null infinity}.
\newblock Class. Quantum Grav. \textbf{26}, 125010 (2012)

\end{thebibliography}

\end{document}